\documentclass[reprint,aps,prl,twocolumn ,groupedaddress,nobibnotes]{revtex4-1}
\usepackage{amssymb}
\usepackage{graphicx}
\usepackage{amsmath}
\begin{document}

\title[]{Large magnetoresistance and Fermi surface study of Sb$_2$Se$_2$Te single crystal}

\author{K. Shrestha$^{1}$}
\email[Corresponding E-mail:]{keshav.shrestha@inl.gov}
\author{V. Marinova$^{2}$}
\author{D. Graf$^{3}$}
\author{B. Lorenz$^{4}$}
\author {C. W. Chu$^{4,5}$}
\affiliation{$^{1}$Idaho National Laboratory,2525 Fremont Ave, Idaho Falls, ID 83401, USA}
\affiliation{$^{2}$Institute of Optical Materials and Technology, Bulgarian Academy of Sciences, Academy G. Bontchev Street 109, Sofia 1113, Bulgaria}
\affiliation{$^{3}$National High Magnetic Field Laboratory, Florida State University, Tallahassee, Florida 32306-4005, USA}
\affiliation{$^{4}$TCSUH and Department of Physics, University of Houston, 3201 Cullen Boulevard, Houston, Texas 77204, USA}
\affiliation{$^{5}$Lawrence Berkeley National Laboratory, 1 Cyclotron Road, Berkeley, California 94720, USA}

\begin{abstract}
  We have studied the magnetotransport properties of a Sb$_2$Se$_2$Te single crystal. Magnetoresistance (MR) is maximum when magnetic field is perpendicular to the sample surface and reaches to a value of 1100\% at $B$=31 T with no sign of saturation. MR shows Shubnikov de Haas (SdH) oscillations above $B$=15 T. The frequency spectrum of SdH oscillations consists of three distinct peaks at $\alpha$=32 T, $\beta$=80 T and $\gamma$=117 T indicating the presence of three Fermi surface pockets. Among these frequencies, $\beta$ is prominent peak in the frequency spectrum of SdH oscillations measured at different tilt angles of the sample with respect to magnetic field. From the angle dependence $\beta$ and Berry phase calculations, we have confirmed the trivial topology of the $\beta$-pocket. The cyclotron masses of charge carriers, obtained by using the Lifshitz-Kosevich formula, are found to be $m^{*}_{\beta}=0.16m_o$ and $m^{*}_{\gamma}=0.63m_o$ for the $\beta$ and $\gamma$ bands respectively. Large MR of Sb$_2$Se$_2$Te is suitable for utilization in electronic instruments such as a computer hard disc, high field magnetic sensors, and memory devices.

\end{abstract}

\pacs{}

\maketitle

\indent
\section{Introduction}
 The recently discovered topological insulators have garnered enormous attention because of their unique surface properties, known as topological surface states or edge states, which arise from the non-trivial topology of the bulk state wave functions in Hilbert space\cite{Hasan:01,Qi:02,Ando:03,Ando1:04,Cava:05}.
 The surface or edge states in topological insulators are important not only for understanding fundamental physics but they also have great technological value for future electronics\cite{Analytis:12}. To exploit the surface states of these materials in future technology, we must first detect and understand their physical properties. Several experimental techniques have been employed to detect topological surface states and understand their properties.
 Angle-resolved photoemission spectroscopy (ARPES) experiments allow one to observe a Dirac cone due to its surface states and a well defined bulk band structure\cite{Hasan:01,Qi:02,Ando:03}. From a technological point of view, transport characteristics of surface states are important. However, the bulk conduction channel interferes with the surface states which makes transport studies of topological surface states challenging\cite{Qu:11,Analytis:12,Eto:13,Cao:14}. Two transport methods, quantum oscillations and weak antilocalization (WAL), have been widely used for studying topological surface states\cite{Taskin:06,Shrestha:15,Bao2012,He2011}. In the presence of magnetic fields, the electrical resistivity shows quantum oscillations, known as Shubnikov-de Haas (SdH) oscillations, due to the quantization of carrier states into Landau levels\cite{KShrestha1,Kittel2006,Ashcroft1976}. The angle dependence of the frequency of SdH oscillations helps to map the cross section of the Fermi surface and its possible origin to either surface or bulk states. Topological materials possess the strong spin-orbit interactions that are often reflected as a cusp or dip in magnetoconductivity at low magnetic fields, known as a weak antilocalization\cite{He2011}$^{,}$ \cite{Chen2014,Shrestha2017}. As a quantum correction to the classical conductivity, a WAL effect could be due to spin-orbit interactions in the surface or bulk states. The scaling of WAL curves with the normal field components provides evidence of the dominance of surface states in magnetoconductivity. Also, a large non-saturating magnetoresistance (MR) and high mobility have been seen in many topological systems, and the observation of these properties has been taken as a signature of the existence of a linear dispersion relation of the surface states\cite{Wang2012,Yan2013}.

 Recently, we have investigated magnetotransport properties of $p$-type metallic Sb$_2$Te$_2$Se single crystal\cite{KShrestha} in fields up to 31 T. From the angle dependence of quantum oscillations and Berry phase calculations, we have confirmed the existence of a 2D Fermi surface in Sb$_2$Te$_2$Se. In the search of a new topological system, we have extended our study to an isostructural compound Sb$_2$Se$_2$Te.

 Sb$_2$Se$_2$Te is a structural analogue of the ternary tetradymite-like\cite{Xu:17} compound Sb$_2$Te$_2$Se with a ``Se" atom in the place of a ``Te" atom. First principles calculation shows that a topological phase transition (from non-trivial to trivial) takes place\cite{Zhang2010} in Sb$_2$(Te$_{1-x}$Se$_{x}$)$_{3}$ while going from x=0 to x=1. Also, band structure calculations\cite{Menshchikova2013} show that Sb$_2$Se$_2$Te is a topologically trivial system with a narrow band gap of  100 meV (direct). Despite these theoretical reports, there are only few transport studies on the Sb$_2$Se$_2$Te compound. Recently, Huang $et$ $al.$\cite{Huang2016} reported a large non-saturating linear MR of Sb$_2$Se$_2$Te that reaches up to 120\% at 9 T. However, due to low field range they could not observe quantum oscillations and study Fermi surface properties. Here, we have studied MR and Fermi surface properties of a Sb$_2$Se$_2$Te single crystal in the extended field range of 31 T. The Fermi surface is three dimensional and has three pockets ($\alpha$, $\beta$ and $\gamma$). Also, we have shown the trivial topology of the $\beta$ pocket using the angle dependence measurements of $\beta$ and Berry phase calculations.
\section{Experimental Procedure}
High quality single crystals of Sb$_2$Se$_2$Te were grown by the modified Bridgman method. The synthesis was done by using stoichiometric quantities of the starting materials Sb, Se, and Te, each with purity of 99.9999\%, mixed in quartz ampoules with diameters of 20 mm and vacuum pumped to 10$^{-6}$ torr. These ampoules were positioned in a Bridgman crystal growth furnace. In the furnace, the ampoules were heated to 700$^{o}$C and homogenized for 48 hours. The crystal growth process was performed with temperature decreasing at 0.3$^{o}$C per hour in the range of 700 $-$ 570 $^{o}$C. The ampoules were further cooled from 570$^{o}$C to room temperature at 10$^{o}$C per hour.\\
\indent Resistivity and Hall measurements were carried out using the ac-transport option of a physical property measurement system (PPMS, Quantum Design). Magnetoresistance measurements were performed at the National High Magnetic Field Laboratory (NHMFL), with fields up to 31 T. Six gold contacts were sputtered on a freshly cleaved crystal face for standard resistivity and Hall measurements. Platinum wires were attached using silver paint. The sample was then mounted on the rotating platform of a standard probe designed at NHMFL. Alternating current of 1 mA was passed through the sample using a Keithley (6221) source meter. The longitudinal and Hall resistances were measured using a lock-in amplifier (SR 830). The sample can be positioned at different angles with respect to the applied magnetic field. The measuring probe is then inserted into a $^3$He Janis cryostat that is mounted on the top of a resistive magnet (31 T). The position of the sample with respect to the applied field was calibrated using a Hall sensor.
\section{Results and Discussion}
\begin{figure}
  \centering
  \includegraphics[width=0.9\linewidth]{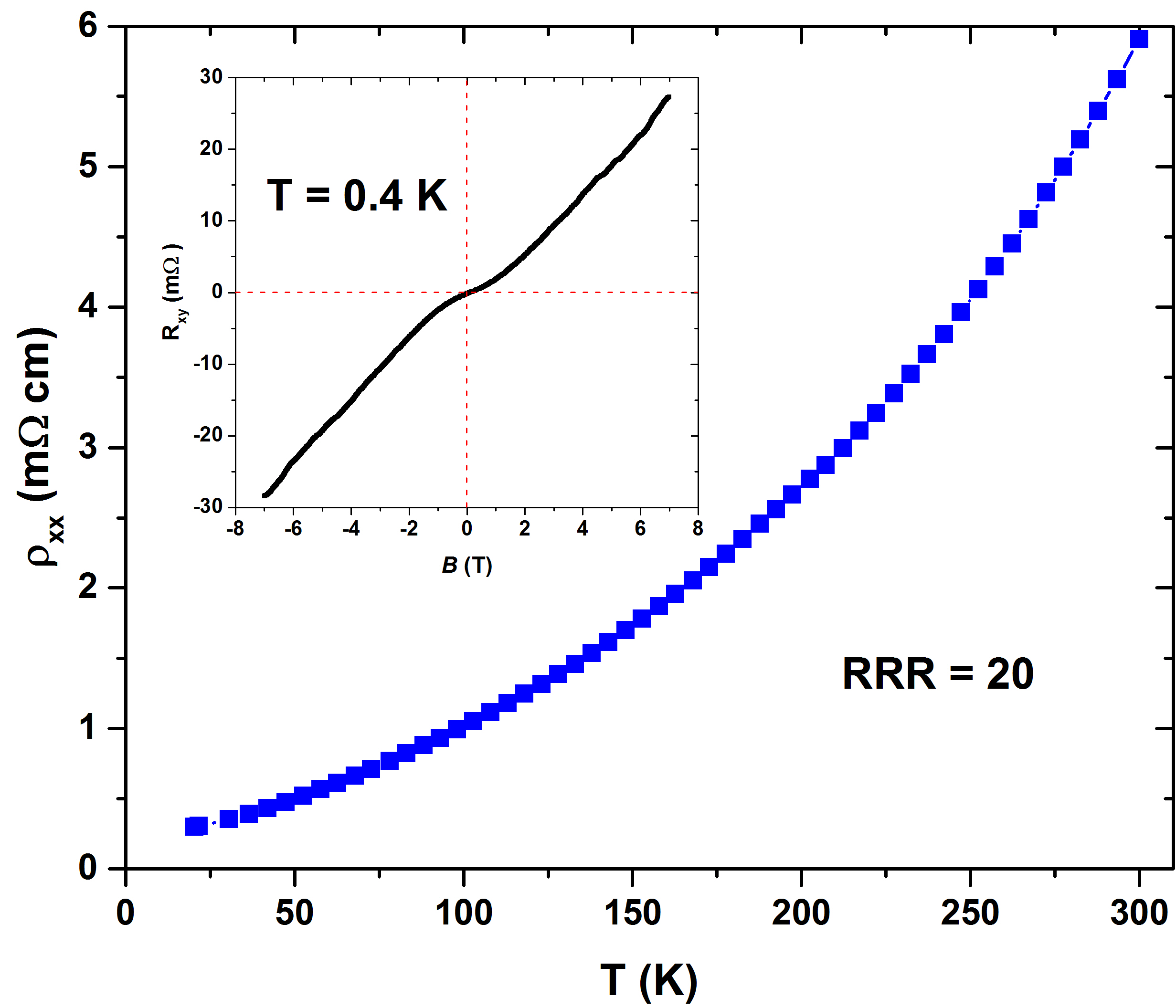}
  \caption{Temperature dependence of resistivity of a Sb$_2$Se$_2$Te single crystal. Inset: Hall resistance measured at $T$=0.4 K. The positive slope of the Hall resistance shows the dominance of the hole-like bulk charge carriers.}\label{Resistivity}
\end{figure}
Figure [1] shows the longitudinal resistivity as a function of temperature for a Sb$_2$Se$_2$Te single crystal. The sample exhibits a metallic behavior below room temperature. The residual resistivity ratio RRR=$\rho_{xx}$(300 K)/$\rho_{xx}$(20 K) = 20 shows the high crystallinity of the single crystal. This value of RRR is nearly 3 times as large as the previous measurement\cite{Huang2016}, RRR = $\rho_{xx}$(300 K)/$\rho_{xx}$(2 K) = 7. Hall measurements were carried out to determine the nature of the bulk charge carriers and their concentration. Non-linear field dependence of $R_{xy}$ near $B$ = 0 suggests the existence of a multiband effect (electron and hole bands)\cite{Qu:11,Ren:21}. However, the overall positive slope of the Hall resistance (see inset, Fig. 1) shows the dominance of the hole-like bulk charge carriers. The bulk carrier concentration is estimated to be 3$\times$10$^{19}$cm$^{-3}$ at 0.4 K.

\begin{figure*}
  \centering
  \includegraphics[width=0.9\linewidth]{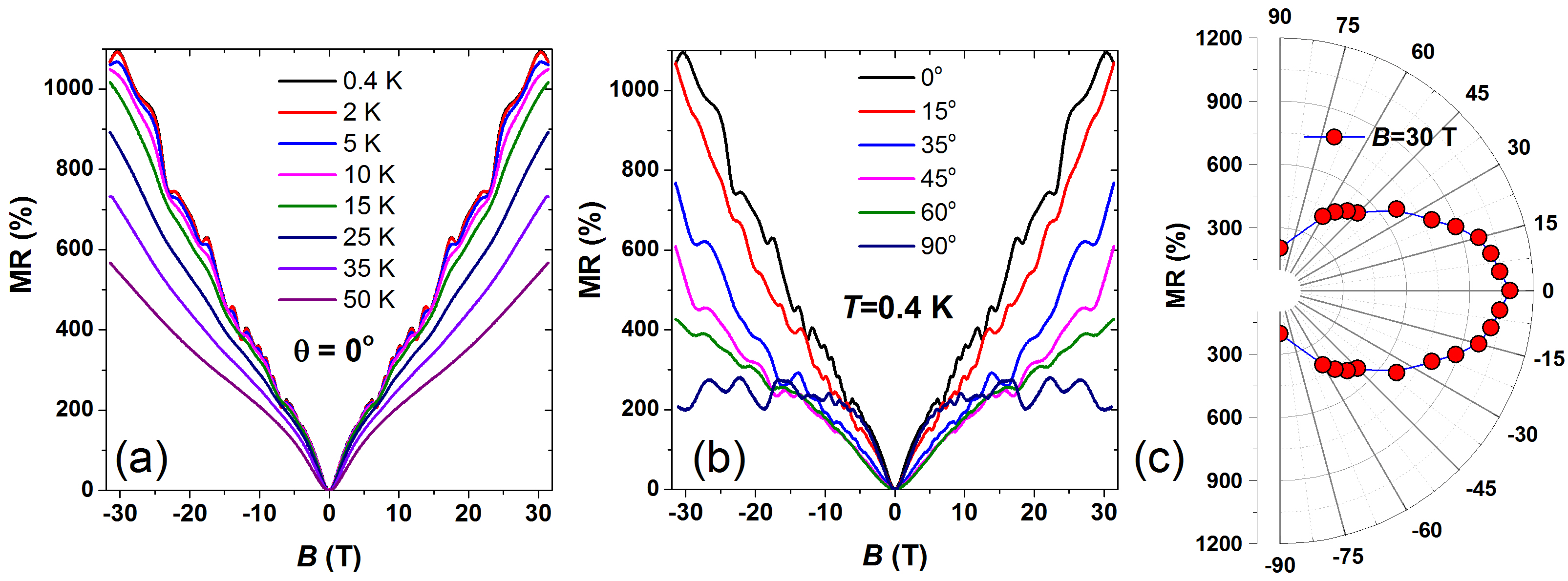}
  \caption{ (a) Temperature dependence of magnetoresistance of Sb$_2$Se$_2$Te single crystal measured at $\theta$=0$^{o}$. (b) Selected magnetoresistance curves at different angles of rotation $\theta$ at $T$=0.4 K. (c) A polar plot of magnetoresistance measured at 30 T. Magnetoresistance is maximum at $\theta$=0$^{o}$ and decreases at higher $\theta$ values.}\label{SdH}
\end{figure*}
 The magnetoresistance of Sb$_2$Se$_2$Te is measured under high magnetic field up to 31 T at NHMFL. Figure [2(a)] shows magnetoresistance expressed in percentage, MR = $\frac{R_{xx}(B)-R_{xx}(0)}{R_{xx}(0)}\times$100\%, measured at different temperature with $\theta$ = 0$^{o}$. Here, $\theta$ is defined as the angle between the magnetic field direction and normal to the sample surface. MR is positive and increases linearly with applied field $B$, and it reaches to 312\% at 9 T. At given field $B$=9 T and $T$=10 K, this MR value is significantly higher than the previous report\cite{Huang2016} of MR=120\%. The large MR response in our sample could be due to better crystallinity i.e. RRR=20 as compared to that of RRR=7 in \cite{Huang2016}. MR decreases gradually with increasing temperature. MR increases linearly with $B$ and reaches $\approx$1100\% at $T$=0.4 K and $B$=31 T. Under same magnetic field and temperature, this MR value is significantly higher than our previous report\cite{KShrestha} of $\approx$98\% increase for Sb$_2$Te$_2$Se single crystal. At $T$=50 K, MR reaches 560\%, which is almost $\frac{1}{2}$ of the value at $T$=0.4 K. A system with large MR usually also possesses high mobility, as observed in many topological and Dirac systems\cite{Shekhar2015,Liang2015}. We have used the simple Drude model [$\mu(T)$=$R_{H}(T)/\rho_{xx}(T)$, where $R_{H}(T)$ is the Hall coefficient at temperature T] to estimate the effective mobility. From our calculations, we found $\mu$$\approx$700 cm$^{-2}$V$^{-1}$s$^{-1}$ at 20 K. It is important to note that MR shows Shubnikov de Haas (SdH) oscillations in the fields above 15 T as a result of the cyclotron motion of the charge carriers in a perpendicular magnetic field. The oscillations are clear at low temperatures and diminished at higher temperatures above $T$=15 K. The frequency of SdH oscillations is proportional to the cross-section area of the Fermi surface. Thus, in order to better understand the shape, size, and dimensionality of the Fermi surface in Sb$_2$Se$_2$Te, we have studied the angle dependence of the quantum oscillations. Figure [2(b)] shows the selected MR curves measured at different $\theta$ values. It is important to note that (1) SdH oscillations are present in all angles from $\theta$=0 to 90$^o$ and (2) the oscillations look complex which could be the mixture of multiple frequencies (will be discussed later). This scenario is different than what we observed in our previous study\cite{KShrestha} on Sb$_2$Te$_2$Se where oscillations possess a single frequency and are diminished above $\theta$=40$^{o}$. The presence of SdH oscillations even at $\theta$=90$^o$ indicates the presence of a 3D Fermi surface in Sb$_2$Se$_2$Te. Also, MR value depends on the tilt angle of the sample with respect to magnetic field. MR is maximum when magnetic field is perpendicular to the sample surface, i.e. at $\theta$=0$^{o}$ and it decreases with increasing $\theta$ as shown in Fig. [2(c)].\\
\begin{figure}
  \centering
  \includegraphics[width=1.0\linewidth]{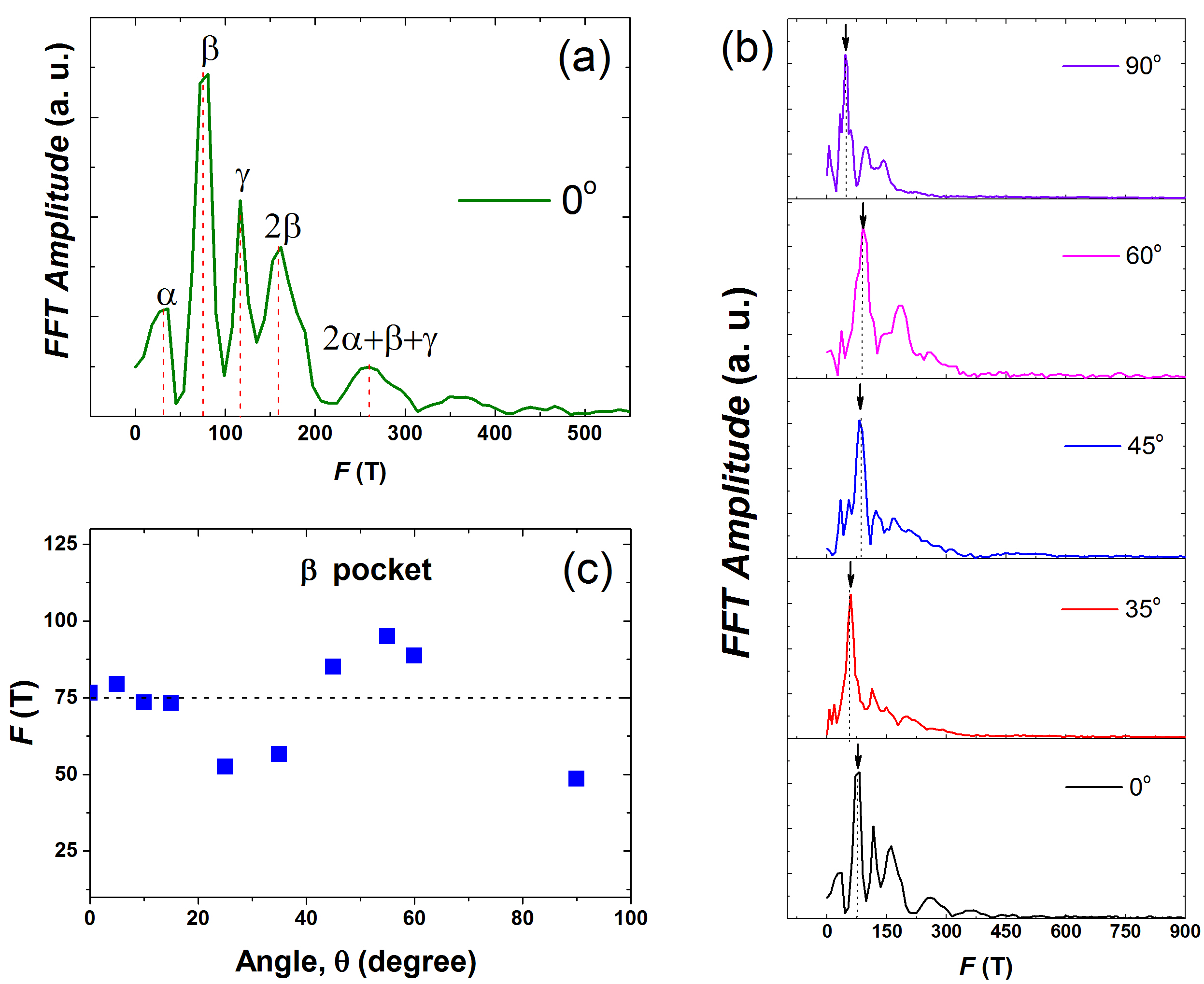}
  \caption{(a) The frequency spectrum of the SdH oscillations at $\theta$=0$^o$. There are three major frequency peaks $\alpha$, $\beta$ and $\gamma$ in the spectrum. The position of frequencies is shown by the dashed lines. (b) Frequency of SdH oscillations measured at selected angles of rotation. (c) Angular dependence of SdH frequency for the $\beta$ band.}\label{FFT}
\end{figure}
\indent The frequency of SdH oscillations shown in Fig. [2(b)] were determined by taking the fast Fourier transform (FFT). Figure [3(a)] shows the FFT spectrum at $\theta$=0$^o$. It consists of three major peaks at 32, 80 and 117 T, denoted as $\alpha$, $\beta$ and $\gamma$. There are two additional peaks, one at 160 T is the second harmonics of $\beta$ and another at 258 T is the sum of 2$\alpha$, $\beta$ and $\gamma$. A frequency of oscillations $f$ is linked to the cross-section of the Fermi surface by Onsagar's relation\cite{Shrestha:15} as $f=(\frac{h}{4\pi e})K^{2}_{F}$, where h is Planck's constant and $K_F$ is the Fermi wave vector. The presence of $\alpha$, $\beta$ and $\gamma$ peaks in Fig. [3(a)] implies the presence of three Fermi pockets in Sb$_2$Se$_2$Te. Fig. [3(b)] shows the FFT of SdH oscillations at selected $\theta$ values. As compared to $\alpha$ and $\gamma$ peaks, the $\beta$ peak is prominent and convenient to track its position at different $\theta$ values, as pointed by black arrows. If quantum oscillations are originated from a 2D Fermi surface, the SdH oscillations or their frequency scale with the normal component of magnetic fields\cite{Ando:03,Ando1:04}. Here, the position of $\beta$ changes with $\theta$ but it does not show any systematic $\theta$ dependence as shown in Fig. [3(c)]. This suggests that the $\beta$ pocket of Fermi surface has a trivial topology.
\begin{figure}
  \centering
  \includegraphics[width=1.0\linewidth]{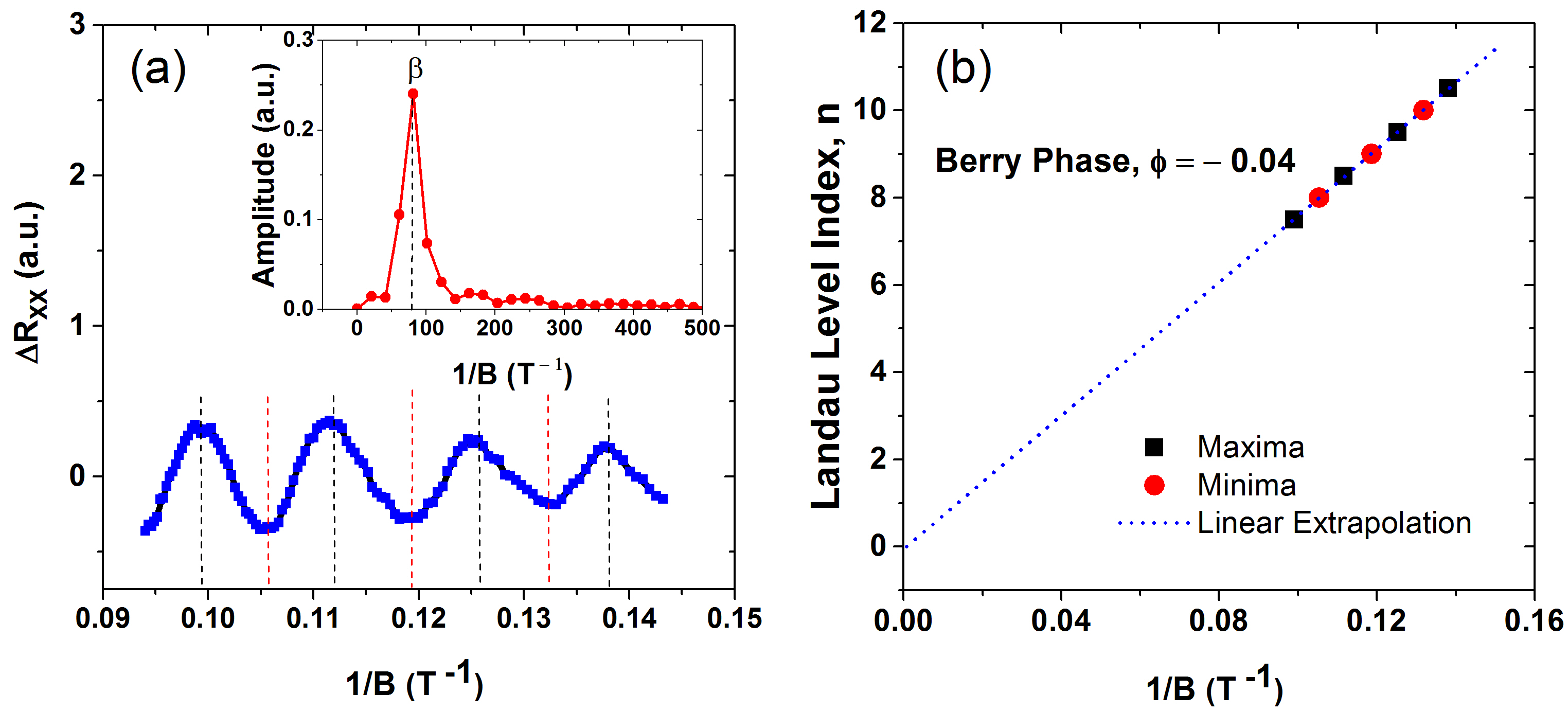}
  \caption{(a) SdH oscillations after subtracting a smooth polynomial background in the field range 7 to 11 T. The positions of maxima and minima are shown by the dashed lines. Inset: FFT spectra of quantum oscillations in 7 to 11 T field range. (b) Landau level fan plot of the $\beta$ pocket. The blue dashed line is the linear extrapolation to zero.}\label{FFT}
\end{figure}

In order to provide further evidence for the bulk states origin of the $\beta$ peak, we have estimated the Berry phase, $\phi$, by using the Landau level (LL) fan plot, which should be $\phi$=0.5 for Dirac particles and 0 for normal fermions. To calculate Berry phase of the $\beta$ oscillations it has to be resolved separately, without the interference from the oscillations of the $\alpha$ and $\gamma$ branches. Figure 4(a) shows the SdH oscillations in magnetic field range (7 $-$ 11) T. In this field range, there exists only one frequency $\beta$ as shown in inset, implying that the contribution from $\alpha$ and $\gamma$ oscillations is completely eliminated. We let 1/$B_{max}$ and 1/$B_{min}$ represent the positions of maxima and minima of the SdH oscillations. The positions of minima and maxima were assigned integer and half integer values respectively to construct the LL fan diagram shown in figure [4(b)]. In the limit 1/$B$ $\rightarrow$0, we have obtained Berry phase, $\phi$=$-$0.04 $\pm$ 0.06. This value is very close to the theoretical value 0 for normal fermions. This further confirms the trivial topology of the $\beta$ pocket.
\begin{figure}
  \centering
  \includegraphics[width=1.0\linewidth]{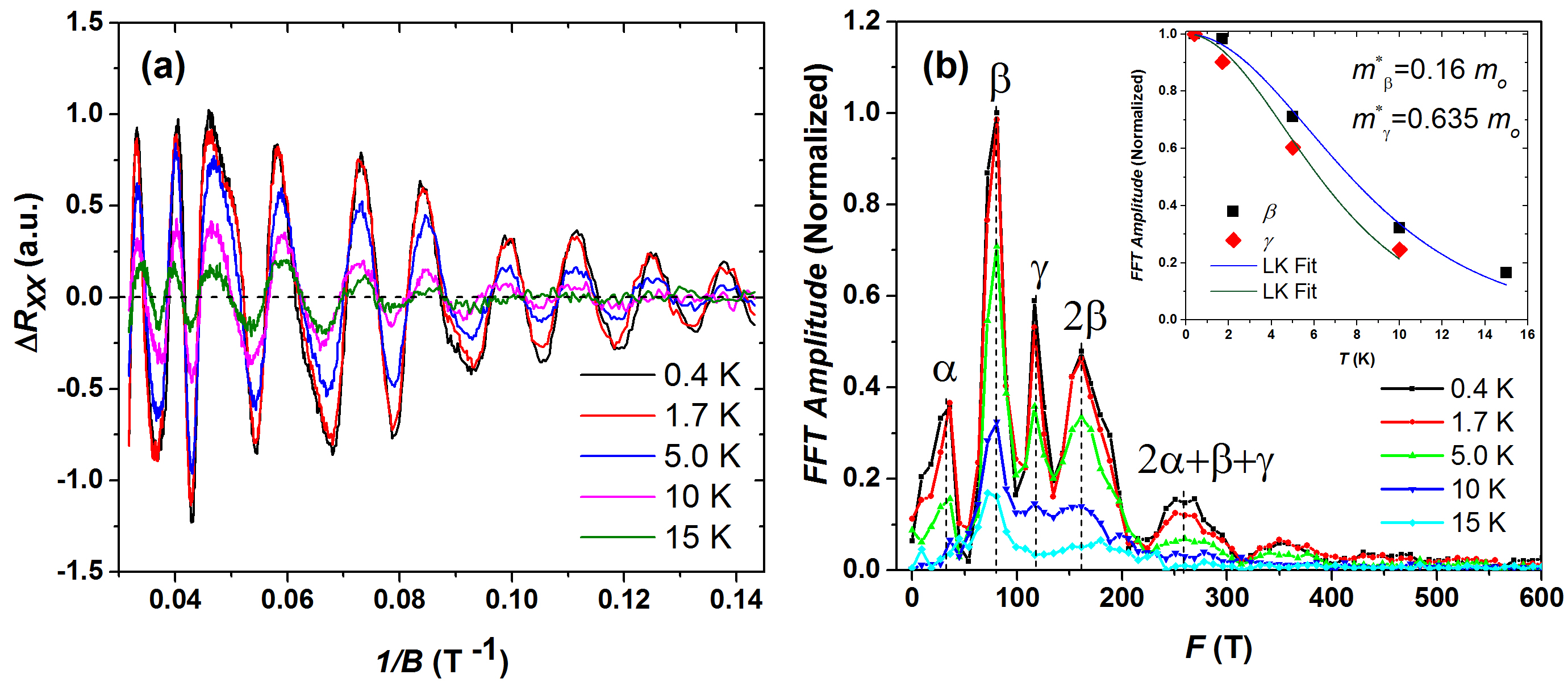}
  \caption{(a) SdH oscillations after subtracting a smooth polynomial background at different temperatures. (b) FFT spectrum of the quantum oscillations shown in (a). Inset: The normalized FFT peak amplitudes for the $\beta$ and $\gamma$ pockets. The solid curves are the LK fittings to the data.}\label{FFT}
\end{figure}

The effective mass $m^{*}$ of charge carriers of Fermi surface pocket can be determined by employing the Lifshitz-Kosevich (LK) formula to temperature dependence data of quantum oscillations. Figure [5(a)] shows SdH oscillations obtained after subtracting a smooth polynomial background at different temperatures. The amplitude of oscillations decreases with increasing temperature. According to LK theory, the oscillation amplitude at fixed magnetic fields can be described by thermal damping factor $R_T$\cite{Qu:11,Ando:03} as
\begin{equation}
R_T = {\lambda(T/B)\over {sinh[\lambda(T/B)]}},
\end{equation}
where
$\lambda(T/B)={2\pi^2k_B\over {\hbar e}} m^{*} {T\over B}$. Due to the presence of multiple oscillation frequencies, it difficult to exactly extract the oscillation amplitude from the raw data in Fig. [5(a)]. That is why, we have taken temperature dependence of the peak height on the FFT spectra of $\Delta R_{xx}$ versus 1/$B$ curves as shown in Fig. [5(b)]. The inset of Fig. [5(b)] shows the LK fitting using Eq.(1) for $\beta$ and $\gamma$ pockets. The parameter $B$ used in LK formula fitting is taken as the average inverse fields of the FFT interval. From the best fit, we have estimated cyclotron masses of $\beta$ and $\gamma$ bands to be $m^{*}_{\beta}=0.16m_o$ and $m^{*}_{\gamma}=0.635m_o$, where $m_o$ is the rest mass of electron. Due to limited temperature dependence data of $\alpha$ peak in the FFT spectrum, we could not use LK formula to determine effective mass of the $\alpha$ pocket.
\section{Summary}
 Here, we have studied magnetoresistance (MR) of a Sb$_2$Se$_2$Te single crystal in magnetic fields up to 31 T. MR is large and reaches up to 1100\% under $T$ = 31 T at $B$=0.4 K and does not show any sign of saturation. MR is maximum when magnetic field is perpendicular to sample surface. In order to study Fermi surface properties of Sb$_2$Se$_2$Te, we have measured Shubnikov de Haas (SdH) oscillations at different tilt angles of sample with respect to magnetic field. Unlike the 2D Fermi surface of Sb$_2$Te$_2$Se compound, Sb$_2$Se$_2$Te has the 3D Fermi surface with three surface pockets ($\alpha$, $\beta$ and $\gamma$). From the angle dependence of frequency and the Berry phase calculations, we have confirmed that the $\beta$ pocket has the trivial topology. Using Lifshitz-Kosevich analyses, we have estimated the cyclotron masses of $\beta$ and $\gamma$ pockets as $m^{*}_{\beta}=0.16m_o$ and $m^{*}_{\gamma}=0.63m_o$.

\section*{acknowledgements}
This work is supported in part by the U.S. Air Force Office
of Scientific Research, the T. L. L. Temple Foundation, the J. J. and R. Moores Endowment, and the State of Texas through the TCSUH. V. Marinova acknowledges support from the Bulgarian Science Fund, project FNI-T-02/26. A portion of this work was performed at NHMFL, which is supported by the NSF co-operative Agreement No. DMR-1157490 and the State of Florida. Work at Idaho National Laboratory is supported by the Department of Energy, Office of Basic Energy Sciences, Materials Sciences, and Engineering Division.

\bibliography{SST}

\end{document}